\documentclass[conference]{IEEEtran}
\IEEEoverridecommandlockouts
\usepackage{cite}

\usepackage{array,multirow}
\usepackage{color}
\usepackage[font={small}]{caption}
\usepackage{tikz}
\usetikzlibrary{calc,matrix}

\ifCLASSINFOpdf
  \graphicspath{{../}}
  \DeclareGraphicsExtensions{.png}
\else
\fi
\usepackage[cmex10]{amsmath}
\usepackage{flushend}
\usepackage{subcaption}

\begin{document}
%
\title{Quantifying the Effect of Matrix Structure on Multithreaded Performance of the SpMV Kernel}

\author{\IEEEauthorblockN{Daniel Kimball, Elizabeth Michel, Paul Keltcher$^\ast$, and Michael~M.~Wolf$^\dagger$}
\IEEEauthorblockA{MIT Lincoln Laboratory\\
Lexington, MA 02420\\
\{daniel.kimball, elizabeth.michel\}@ll.mit.edu, paul@keltcher.com, mmwolf@sandia.gov}
\thanks{$^\ast$Now with Advanced Micro Devices.}
\thanks{$^\dagger$Now with Sandia National Laboratories.}
\thanks{This work was sponsored by Defense Advanced Research Projects Agency (DARPA) under Air Force contract FA8721-05-C-0002. The views expressed are those of the authors and do not reflect the official policy or position of the Department of Defense or the United States Government. This document is Approved for Public Release. Distribution Unlimited.}
}


%

\IEEEoverridecommandlockouts
\IEEEpubid{\makebox[\columnwidth]{978-1-4799-6233-4/14/\$31.00~\copyright2014
IEEE \hfill} \hspace{\columnsep}\makebox[\columnwidth]{ }}

\maketitle

\begin{abstract}

Sparse matrix-vector multiplication (SpMV) is the core operation in many common network and graph analytics, but poor performance of the SpMV kernel handicaps these applications. This work quantifies the effect of matrix structure on SpMV performance, using Intel's VTune tool for the Sandy Bridge architecture. Two types of sparse matrices are considered: finite difference (FD) matrices, which are structured, and R-MAT matrices, which are unstructured. Analysis of cache behavior and prefetcher activity reveals that the SpMV kernel performs far worse with R-MAT matrices than with FD matrices, due to the difference in matrix structure. To address the problems caused by unstructured matrices, novel architecture improvements are proposed.


\end{abstract}



%
\IEEEpeerreviewmaketitle

\section{Introduction}


Unstructured, sparse matrices arise in many common applications such
as network analysis, linear dynamical systems, Markov chains, and
eigendecomposition. Performance of these applications depends on the
efficiency of sparse linear algebra operations on 
matrices. Sparse matrix-vector multiplication (SpMV) is the foundation
for many of these operations. For example, SpMV is an essential
kernel in signal processing for graph applications
, where the
computation of principal components (obtained through the use of
eigensolvers) is a key step in finding anomalies in network
data~\cite{spg}. In this and many other applications, SpMV dominates
the runtime, so performance is limited by SpMV efficiency.

%

In previous work, it was shown that the SpMV kernel performs one to
three orders of magnitude worse in GOPS/Watt than its dense
counterpart \cite{PAKCK_HPEC2013}. Dense matrix-vector multiplication
is highly optimized and makes good use of modern architectural
features~\cite{goto:2008}. Sparse matrix-vector multiplication (SpMV)
with structured, sparse matrices benefits from many of the same
features. For SpMV on unstructured, sparse matrices,
however, irregularity of data access causes a high amount of traffic
within the memory hierarchy.  The architectural features of processors
such as caches and prefetchers do not improve the performance of the SpMV
kernel significantly for these matrices.

This work quantifies the performance of the SpMV kernel for
unstructured sparse matrices, using data on structured sparse matrices
for comparison. In Section~\ref{sec:background}, the structure of the
matrices and the architecture used are detailed, as well as the
relationship between the two. Section~\ref{sec:methodology} explains
how the data was collected and how it is used to measure the
performance of various components of the architecture, the caches, and the prefetcher. Results are
presented in Section~\ref{sec:results}. Finally,
Section~\ref{sec:conclusion} discusses the implications of sparse
matrix structure and proposes architectural innovations to improve SpMV
performance.





\section{Background}
\label{sec:background}
\subsection{Matrices}


R-MAT matrices are unstructured and model the type of graphs that
arise frequently in network applications~\cite{rmat}. R-MAT matrices
are constructed to approximate realistic network models using a power
law distribution. This construction skews the distribution of nonzeros
in the matrix. To avoid problems of load balancing across threads, the
rows and columns of the R-MAT matrices are permuted randomly. This
does not change the results of the multiplication and eliminates a
confounding element in comparisons to performance with other
matrices. These R-MAT matrices are generated to represent a network
with an average of eight connections per node, giving an average of
eight nonzeros per row.

Finite difference (FD) matrices are also sparse, but have a more
regular structure than R-MAT matrices. The FD matrices are generated using a
two-dimensional, 9-point stencil~\cite{Saad:2003}. The matrices have three diagonal
bands of three nonzero elements each, giving nine nonzeros per row.

Both types of sparse matrices are stored efficiently in compressed
sparse row (CSR) format~\cite{1973sparse}. The CSR format consists of
three dense arrays: an array of nonzero values, an array of
column indices for each nonzero, and an array of indices to the start of each
new row in the values array.  The total number of elements in these
arrays, for a matrix with $n$ rows and $m$ nonzeros, is $2m+n+1$.

The CSR format stores sparse matrices compactly, which
allows sequential access to the nonzero entries of the matrix.
Matrix accesses, however, are not the only source of memory requests.
To perform matrix-vector
multiplication, each nonzero element (found in some column $j$) is
multiplied by the $j^{th}$ element of an input vector $\vec{x}$. 
For sparse matrices, the order of accesses to $\vec{x}$ is determined
by the location of the nonzeros in the matrix. The kernel's ability to access the
needed elements of $\vec{x}$ determines the performance of
matrix-vector multiplication.

\subsection{Intel Sandy Bridge Architecture}

\begin{figure}[htb]
  \centering
 \includegraphics[scale=0.47]{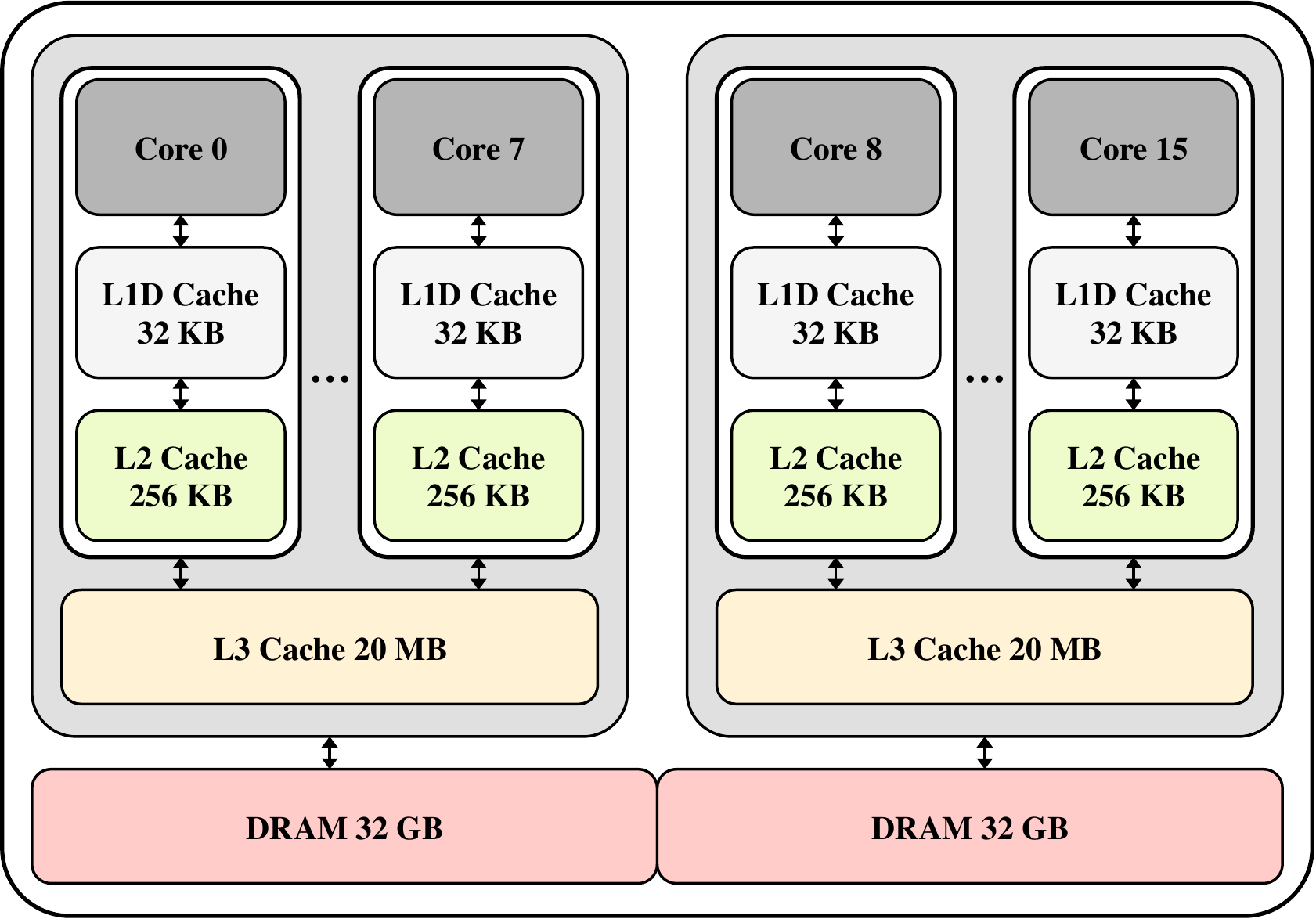}
  \caption{Sandy Bridge Architecture.}
  \label{fig:SandyBridge}
\end{figure}

The SpMV performance measurements were
conducted on Intel Sandy Bridge, a multi-core, non-uniform memory
access architecture. The Sandy Bridge memory hierarchy has two levels
of on-core cache (L1D and L2), a shared on-chip L3 cache, and shared
off-chip DRAM, as shown in Figure~\ref{fig:SandyBridge}. This
architecture has a memory prefetcher, which preemptively loads data
into the L2 cache from the L3 cache (or from DRAM, if the data is not
found in L3) to speed up future access to this data. If the link to
DRAM is too congested with demand requests from the SpMV kernel,
the prefetcher will not turn on~\cite{IntelOpt}.

This architecture is commonly found in laptop, desktop, workstation
and server computers. Sandy Bridge processors are also used in
Amazon's Elastic Compute Cloud and Google's Compute Engine. The Intel
Sandy Bridge architecture was selected because of its widespread use,
its multi-core processor, and the availability of tools to access
various hardware counters in the processor for tracking data such as
cache hits and instruction counts.

For this study, performance data was obtained on a Sandy Bridge
system containing two Intel Xeon E5-2690 processors running at 2.9
GHz, with 32 GB of shared DRAM for each processor.  Each processor has
20MB of on-chip shared L3 cache and contains 8 cores. Each core has
32KB of on-core L1D cache and 256 KB of on-core L2 cache~\cite{IntelXeon}.

\subsection{Cache Utility}

The performance of SpMV depends greatly on the
architecture's ability to utilize cache efficiently. The
multiplication operation uses three data structures that must be stored in
memory: a matrix, an input vector ($\vec{x}$), and a
vector for the solution. When the problem size is small enough that
all three data structures fit in cache, every memory request is satisfied on-chip (cache hit). When the problem size is large, the processors must
access DRAM to obtain data that has not yet been pulled into
cache. Accesses to DRAM are far more costly than accesses to cache and
hinder performance.

The matrices tested have sizes ranging from $8\times2^{11} = 16,384$
nonzeros to $9\times2^{26} = 603,979,776$ nonzeros.\footnote{The R-MAT
  matrix sizes range from $2^{11}$ to $2^{26}$ rows, times 8 nonzeros
  per row. The FD matrices have the same range of rows, times 9
  nonzeros per row.}  The smallest problems fit entirely in the L2
cache, while the largest barely fit in
DRAM. Table~\ref{table:cacheSizes} shows the maximum number of
nonzeros a matrix may have and still fit within each cache level.

\begin{table}[htb]
  \centering
  \renewcommand{\arraystretch}{1.2}
  \begin{tabular}{|r|c|r||r|r|r|} \cline{2-5}
    \multicolumn{1}{r|}{}     & Level & Size  & FD        & R-MAT     \\ \hline
    \multirow{2}{*}{Serial}   & L2    & 256KB &    18,432 &    18,078 \\
                              & L3    & 20MB  & 1,474,560 & 1,446,311 \\ \hline
    \multirow{2}{*}{Parallel} & L2    & 4MB   &   294,912 &   289,262 \\
                              & L3    & 40MB  & 2,949,120 & 2,892,623 \\ \hline
  \end{tabular}
  \caption{Maximum number of nonzeros a matrix can contain, such that
    the problem fits in the given cache level. Data shown for serial execution
    (1 thread, 1 L2 cache, 1 L3 cache) and parallel execution (16 threads, 16 L2
    caches, 2 L3 caches).}
  \label{table:cacheSizes}
\end{table}

For both FD and R-MAT matrices, the CSR matrix data is straightforward
for the system architecture to load. The elements of all three arrays
of the CSR format are stored and accessed sequentially, allowing each
memory request to retrieve an entire cache line of useful data. This
spatial locality enables the L2 prefetcher to anticipate data needs
correctly and retrieve useful data from lower memory in advance. The challenge arises
when accessing the elements of $\vec{x}$.

Caching data from $\vec{x}$ is much easier for matrix-vector
multiplication on structured matrices than on unstructured
matrices. First, consider the ideal case of a dense matrix. In dense
matrix-vector multiplication, elements of $\vec{x}$ are accessed
sequentially (since every element in each row is treated as a
nonzero), so portions of $\vec{x}$ can be fetched into the cache in
advance for efficient access. The pattern of accesses to $\vec{x}$ is
not so simple for SpMV with sparse matrices.

\begin{figure}[htb]
\centering
\begin{tabular}{cm{.11in}c}
\begin{tikzpicture}[baseline=(current bounding box.east)]
\matrix[matrix of math nodes,inner sep=0pt,row sep=.7em,column sep=.7em, 
  left delimiter=[,right delimiter={]}, outer sep=1pt] (M)
{
      1 & 1 & 0 & 0 & 0 & 0 & 0 & 0 &\\
      1 & 1 & 1 & 0 & 0 & 0 & 0 & 0 &\\
      0 & {\bf \color{red}A} & {\bf \color{red}A} & {\bf \color{red}A} & 0 & 0 & 0 & 0 & \\
      0 & 0 & 1 & 1 & 1 & 0 & 0 & 0 & \cdots \\
      0 & 0 & 0 & 1 & 1 & {\bf \color{blue}B} & 0 & 0 & \\
      0 & 0 & 0 & 0 & 1 & {\bf \color{blue}B} & 1 & 0 &\\
      0 & 0 & 0 & 0 & 0 & {\bf \color{blue}B} & 1 & 1 &\\
      0 & 0 & 0 & 0 & 0 & 0 & 1 & 1 & \\
      & & & \raisebox{.5em}[1.4em][\depth]{\vdots} & & & & & \raisebox{.5em}[1.4em][\depth]{$\ddots$}\\
}
;


\draw[->] (M-3-2.east) -- (M-3-3.west);
\draw[->] (M-3-3.east) -- (M-3-4.west);


\draw[->] (M-5-6.south west) -- (M-6-5.north east);

\draw[->] (M-6-5.east) -- (M-6-6.west);
\draw[->] (M-6-6.east) -- (M-6-7.west);
\draw[->] (M-6-7.south west) -- (M-7-6.north east);


\end{tikzpicture}
&
\begin{tikzpicture}
$\times$
\end{tikzpicture}
&
\begin{tikzpicture}[baseline=(current bounding box.east)]
\matrix[matrix of math nodes,inner sep=0pt,row sep=.7em,column sep=.9em, 
  left delimiter=[,right delimiter={]}, outer sep=1pt] (X)
{
      1\\
      {\bf \color{red}A}\\
      {\bf \color{red}A}\\
      {\bf \color{red}A}\\
      1\\
      {\bf \color{blue}B}\\
      1\\
      1\\
      \raisebox{.5em}[1.4em][\depth]{\vdots}\\
}
;

\end{tikzpicture} \\
Finite difference matrix && $\vec{x}$
\end{tabular}
  \caption{Part of an FD matrix is multiplied by an input vector
    $\vec{x}$, showing how the pattern of nonzeros in the matrix
    determines sequential accesses (red As) and repeated accesses
    (blue Bs) to the elements in $\vec{x}$. FD matrices have three
    diagonal bands of three nonzeros each. One of the bands is shown
    here.}
  \label{fig:MatrixExample}
\end{figure}

For FD matrices, accesses to elements of $\vec{x}$ have strong spatial
and temporal locality. In a given row, there are three sets of three
adjacent nonzero elements. For each set, the SpMV kernel requires
three adjacent elements of $\vec{x}$, as shown in
Figure~\ref{fig:MatrixExample} by red letter As. This sequential
access pattern allows the prefetcher to anticipate the needs of the
kernel successfully and load the appropriate elements of $\vec{x}$
into the L2 cache. Since memory accesses are performed in units of a
cache line, additional elements of $\vec{x}$ are loaded and will
be in cache when needed. Accesses to $\vec{x}$ also exhibit temporal
locality: after an element of $\vec{x}$ is used the first time, it gets
used again during the multiplications of the next two rows of the
matrix, since they have nonzeros in the same column (see
Figure~\ref{fig:MatrixExample}, blue letter Bs). Even if the kernel
were to miss in cache for the first usage of an element of $\vec{x}$,
it would hit for the next two uses immediately after. These properties
of FD matrix structure and the resultant accesses to $\vec{x}$ allow
the SpMV kernel to use cache effectively.

For R-MAT matrices, the lack of structure forces the SpMV kernel to
access elements of $\vec{x}$ at random. Without spatial or temporal
locality in accesses to $\vec{x}$, the prefetcher fails to predict
data needs; therefore, useful data is rarely found in cache. This forces the
SpMV kernel to make many more demand requests to DRAM, slowing the
computation and damaging performance.

\section{Methodology}
\label{sec:methodology}
\subsection{Intel VTune Amplifier}

Performance data is measured precisely with Intel's VTune Amplifier XE tool. VTune records activity on various
components of the processor. This study concentrates on cache-related metrics. VTune collects hits and misses for each level of
cache, as well as remote versus local DRAM
accesses. It also captures high-level metrics such as the
number of instructions executed and the number of cycles used. The
most powerful functionality of VTune is the ability to collect data on specific
functions, threads, or even line numbers. In this work, the data is collected exclusively from the SpMV operation. To ensure   
the SpMV operation uses enough resources for VTune to track it accurately, the SpMV kernel is run many times and the result is averaged over the number of
runs. The number of times SpMV is run per matrix is:
$$\mbox{Number of Runs} = \frac{2^{33}}{\mbox{\# of non-zeros}}$$
This function has the property that the amount of computational work performed is constant and independent of the matrix size. This allows for comparison of the counters
across matrix sizes without normalizing. Additionally, the SpMV program is run without VTune to measure the runtime of the SpMV kernel, to ensure that gathering the metrics does not slow the runtime measurement.

\subsection{Metrics}

From the VTune counters\footnote{
The VTune counters are renamed above for readability. Here is the correspondence between the names above and the identifiers used in VTune:\\
{\scriptsize \centering
\setlength{\tabcolsep}{2pt}
\begin{tabular}{ll}
\hspace{10pt}L2 Demand Misses       & MEM\_LOAD\_UOPS\_RETIRED.L2\_MISS\\
\hspace{10pt}L3 Demand Misses       & MEM\_LOAD\_UOPS\_RETIRED.LLC\_MISS\\
\hspace{10pt}Prefetcher L2 Misses   & L2\_RQSTS.PF\_MISS\\
\hspace{10pt}L2 Cycles Stalled      & CYCLE\_ACTIVITY.STALL\_CYCLES\_L2\_PENDING\\
\hspace{10pt}Number of Instructions & INST\_RETIRED.ANY\\
\hspace{10pt}Total Number of Cycles & CPU\_CLK\_UNHALTED.THREAD\\
\end{tabular}}
}, five compound metrics are computed to capture
concisely the relevant performance information. These five metrics are
L2 Miss Rate, L3 Miss Rate, Prefetch Miss Rate, L2 Stall Cycles, and GFLOPS.

L2 Miss Rate is a measure of how often the process misses the L2 cache
for demand requests.  A high cache miss rate indicates poor utilization of
the cache. The L2 Miss Rate is computed as:
\begin{equation} \label{eq:L2MR}
\mbox{L2 Miss Rate}=10^3\times\frac{\mbox{L2 Demand Misses}}{\mbox{Number of Instructions}}
\end{equation}\vspace{0pt}

L3 Miss Rate is the equivalent metric for the L3 cache. A high cache
miss rate (in the L2 or L3 caches) causes the system to waste time
waiting for the request to complete.\footnote{Data on the L1D cache
  was not included because L1D miss rate would not affect L2 Stall
  Cycles, one of the primary metrics used to measure performance} The L3 Miss Rate is computed as:
\begin{equation} \label{eq:L3MR}
\mbox{L3 Miss Rate}=10^3\times\frac{\mbox{L3 Demand Misses}}{\mbox{Number of Instructions}}
\end{equation}\vspace{0pt}

 Prefetch Miss Rate measures how often the prefetcher loads data into
 the L2 cache. A high Prefetch Miss Rate indicates prefetcher success
 in speeding up the SpMV operation. It is important to note that this
 is contrary to the L2 and L3 Miss Rates in that a higher Prefetch
 Miss Rate implies better performance. Prefetch Miss Rate is computed
 as:
\begin{equation} \label{eq:PFMR}
\mbox{Prefetch Miss Rate}=10^3\times\frac{\mbox{Prefetcher L2 Misses}}{\mbox{Number of Instructions}}
\end{equation}\vspace{0pt}

L2 Stall Cycles is the percent of total cycles spent waiting for data
in the L2 cache. L2 Stall Cycles includes not only L2 misses but also
subsequent L3 accesses and fetches from DRAM. These are the cycles that
would have been saved, had the data been present in L2. L2 Stall Cycles
is computed as:
\begin{equation} \label{eq:L2SC}
\mbox{L2 Stall Cycles}=\frac{\mbox{L2 Cycles Stalled}}{\mbox{Total Number of Cycles}}
\end{equation}\vspace{0pt}

GFLOPS shows performance in billion floating-point operations per
second. In SpMV, each non-zero element is involved in one
multiplication and one addition, so the number of floating point
operations is twice the number of nonzeros.  GFLOPS is inversely
proportional to the runtime, so faster runtimes correspond to higher
GFLOPS.  GFLOPS is computed as:
\begin{equation} \label{eq:GFLOPS}
\mbox{GFLOPS}=\frac{1}{10^9}\times\frac{2\times\mbox{Number of Nonzeros}}{\mbox{SpMV runtime}}
\end{equation}\vspace{0pt}




\section{Results}
\label{sec:results}
\begin{figure*}[t]
\centering
	\begin{subfigure}[b]{0.98\columnwidth}
        	\centering       
        	\includegraphics[width=1\textwidth]{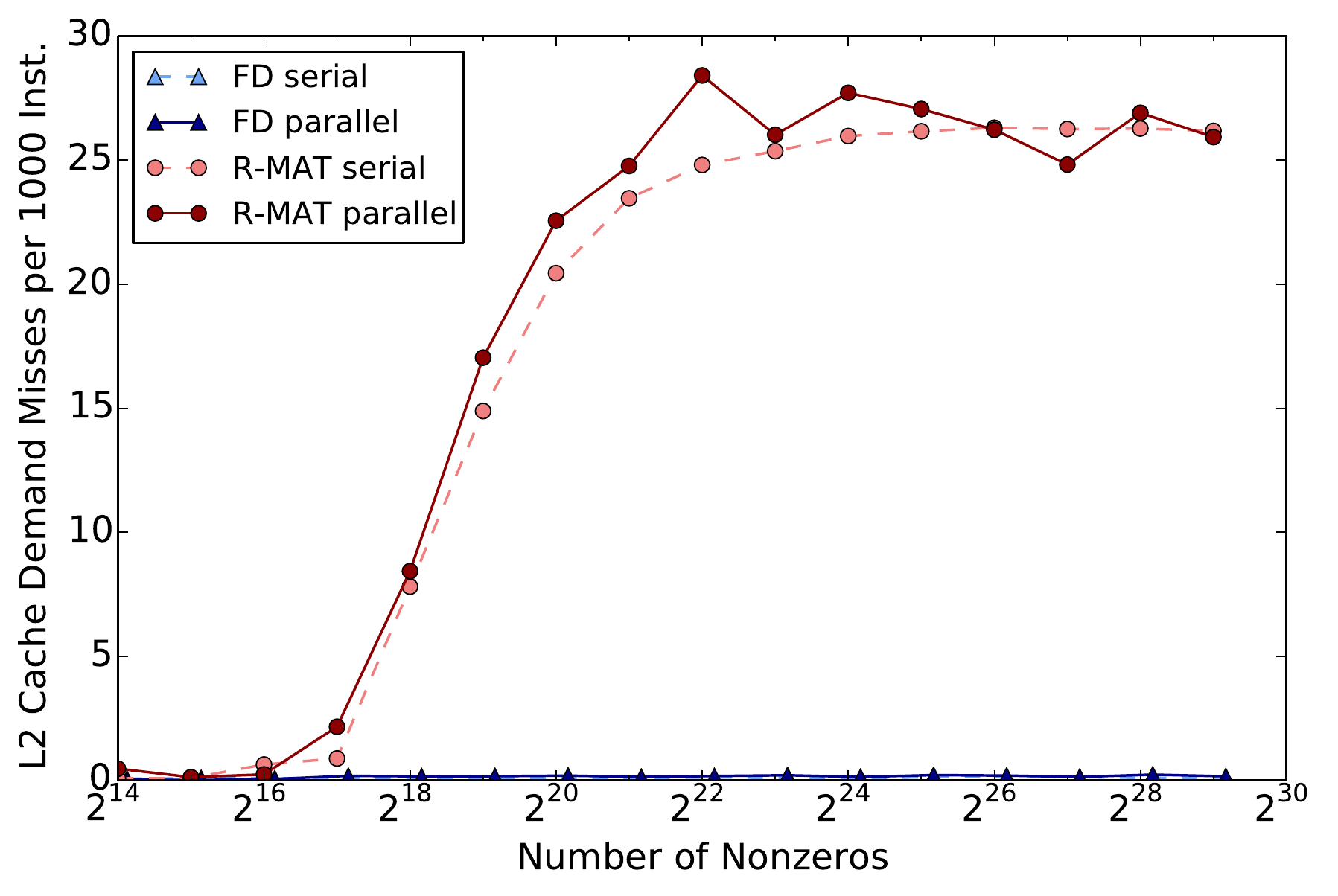} 
        	\caption{L2 Miss Rate.}
        	\label{fig:L2MR}
	\end{subfigure}
	\hspace{\columnsep}
	\begin{subfigure}[b]{0.98\columnwidth}
        	\centering       
        	\includegraphics[width=1\textwidth]{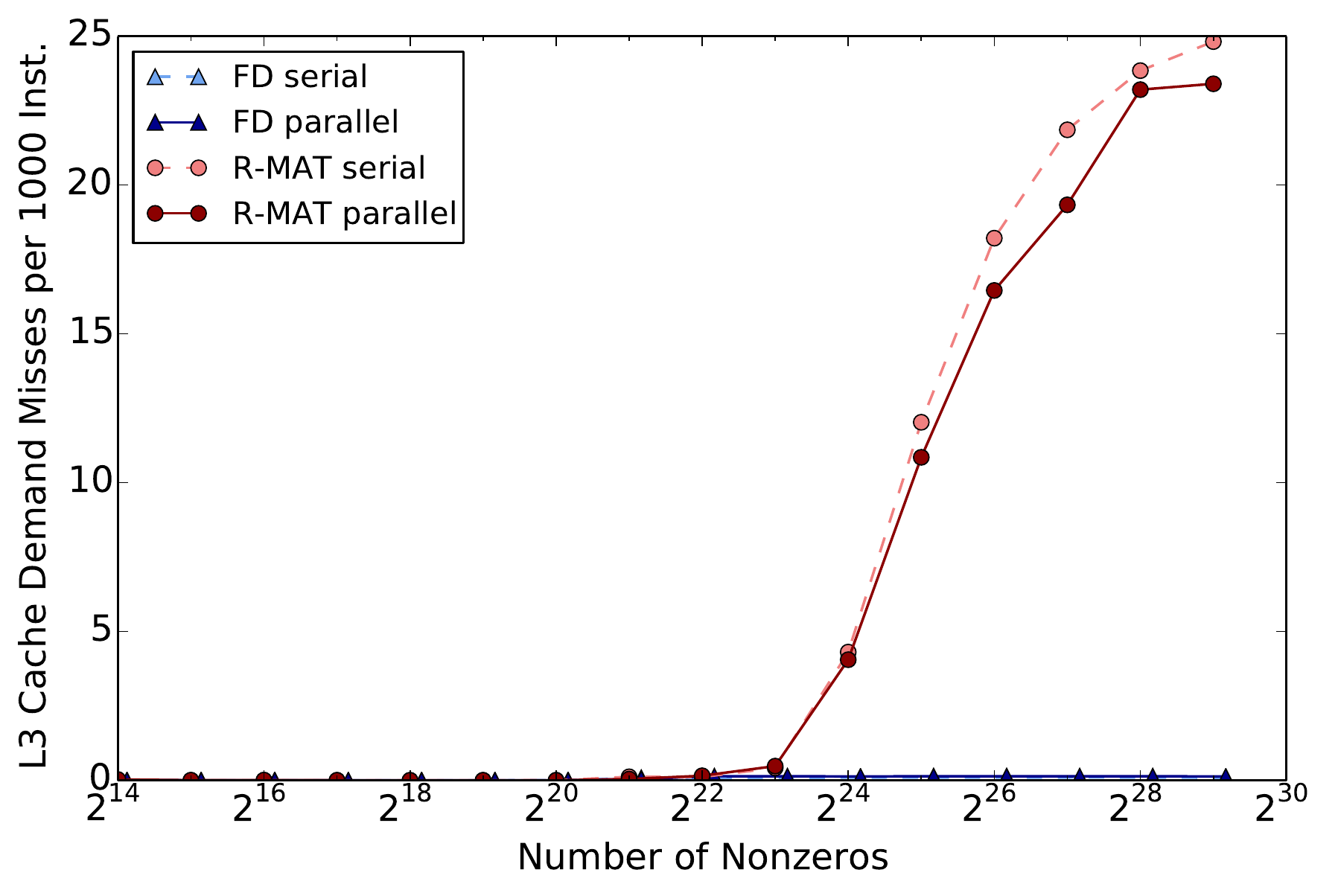} 
        	\caption{L3 Miss Rate.}
       		\label{fig:L3MR}
	\end{subfigure}
        \caption{Cache miss rates of the SpMV kernel on FD matrices (blue) and R-MAT
          matrices (red) in serial and in parallel. Cache miss rates for FD matrices 
          are around zero, but miss rates for R-MAT matrices grow with increasing 
          matrix size}
        \label{fig:CacheMR}
\end{figure*}

Performance data on the SpMV kernel is presented for 1, 2, 4, 8, and
16 threads. Data is not shown for 32 threads, as the results are confounded by
hyperthreading.
\footnote{The system has 16 physical cores and uses hyperthreading to
  simulate an additional 16 cores~\cite{IntelXeon}. Hyperthreading
  does not duplicate the resources essential to the SpMV kernel, so
  there is no performance gain from attempting to use more threads
  than physical cores. In fact, simulating additional cores only
  fragments resources and increases congestion. }
For cache metrics,
two representative cases are presented for simplicity. The serial case
(1 thread) is presented to show performance without the congestion of
additional threads. Data for 16 threads is used as the parallel case,
to show maximal utilization of all processor resources by the
kernel. Performance is quantified in terms of L2 Miss Rate, L3
Miss Rate, Prefetcher Miss Rate, L2 Stall Cycles, and GFLOPS.

\subsection{L2 Miss Rate}

Figure~\ref{fig:L2MR} shows the L2 Miss Rate (Eq.~\ref{eq:L2MR})
of the SpMV kernel for FD and R-MAT matrices. For the FD matrices, the
SpMV kernel maintains a low L2 Miss Rate of about 0.1 misses per
thousand instructions across all matrix sizes. For the R-MAT matrices,
as the problem size exceeds the capacity of the L2 cache, L2 Miss Rate
increases dramatically, reaching a plateau around 26 misses per thousand
instructions. This discrepancy is caused by the difference in
structure between the two matrix types.

The regular structure of the FD matrices results in some sequential
accesses to $\vec{x}$ by the SpMV kernel and allows reuse of some data
in the L2 cache. The consequence of this regular access pattern is
that the needed data is regularly present in the L2 cache, producing a
low (good) L2 Miss Rate.

For R-MAT matrices, their unstructured composition generates random
accesses to $\vec{x}$. These erratic accesses to $\vec{x}$ hinder the
architecture's ability to keep relevant data in the L2 cache, causing
a high (bad) L2 Miss Rate.

The L2 Miss Rate differs little between the SpMV kernel running in
serial and in parallel. Although running in parallel provides multiple
L2 caches, the increased net capacity does not increase L2 cache
capacity for any individual core. The performance of each core depends
only on the data present within its own L2 cache, not on the data
within the L2 caches of other cores. The performance of each core for
a given matrix size, therefore, remains the same as in the serial
case.

The L2 Miss Rates level out when only a trivial portion of the problem
fits in the L2 cache. For these large problem sizes, data accesses to
elements of $\vec{x}$ by the SpMV kernel settle into a pattern of
where the data is found. For FD matrices, the pattern is that the
hardware prefetcher successfully anticipates data requirements and
loads the correct data into the L2 cache. For R-MAT matrices, the
prefetcher fails to predict data requirements (due do the random R-MAT
structure), so the data access pattern is as follows: look for the data in the
L2 cache, miss, and then retrieve the data from a more remote location in
memory.

\begin{figure*}[t]
\centering
 \begin{minipage}{.98\columnwidth}
        \centering       
        \includegraphics[width=1\textwidth]{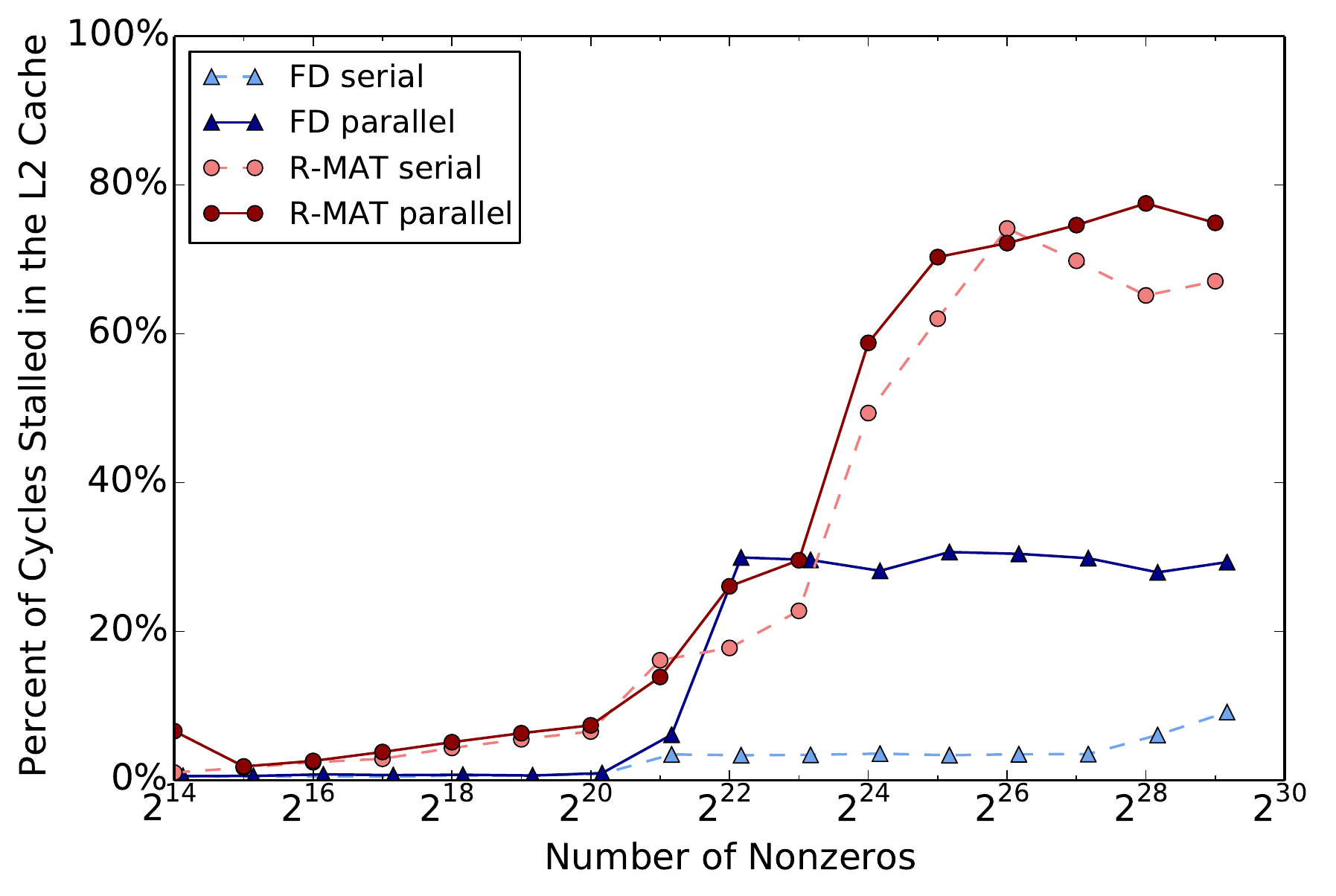} 
        \caption{L2 Stall Cycles of the SpMV kernel on FD (blue) and R-MAT (red)
          matrices in serial and in parallel.}
        \label{fig:L2SC}
 \end{minipage}
 \hspace{\columnsep}
 \begin{minipage}{.98\columnwidth}
        \centering
        \includegraphics[width=1\textwidth]{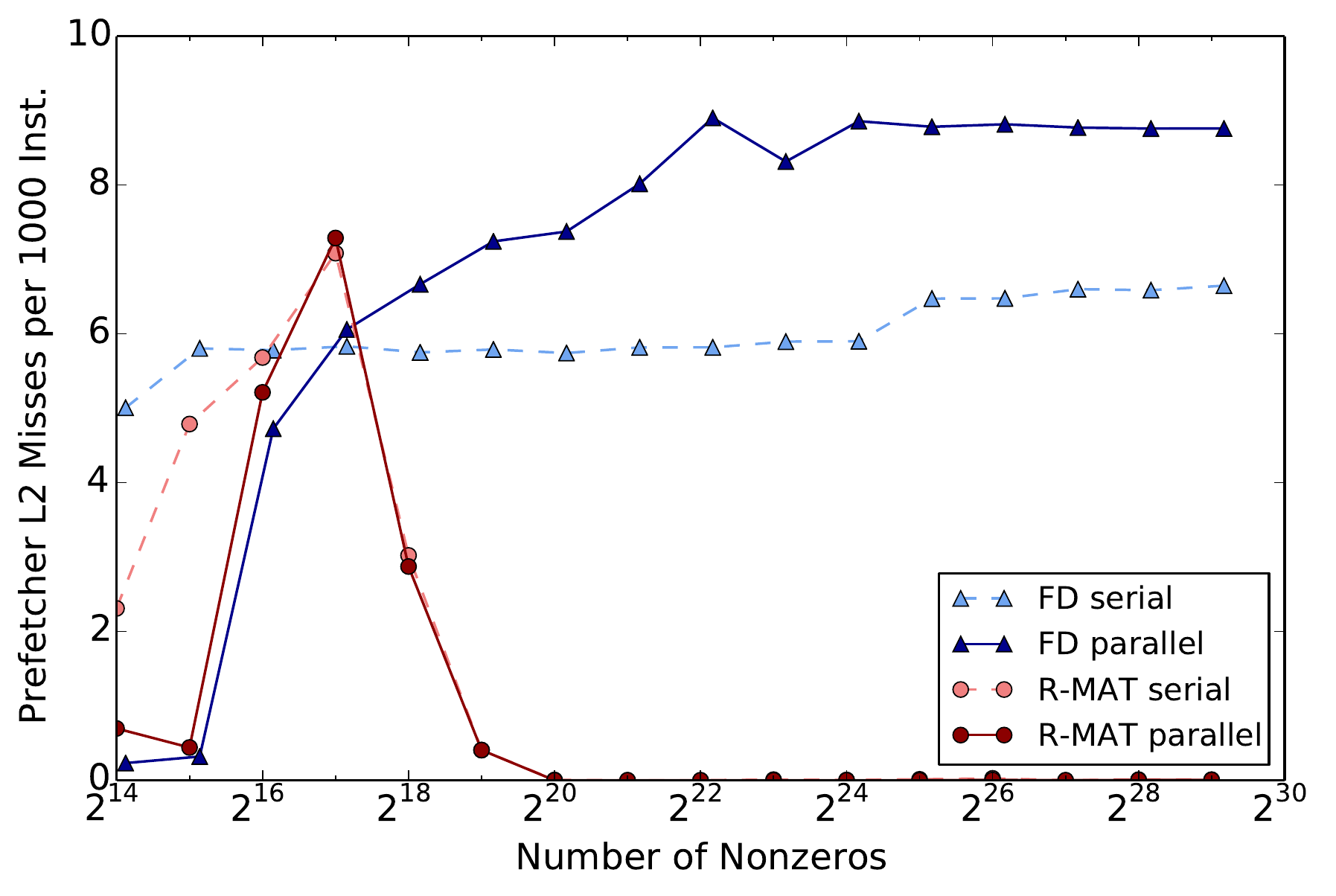}
        \caption{Prefetch Miss Rate of the SpMV kernel on FD matrices (blue) and
          R-MAT matrices (red) in serial and in parallel.}
        \label{fig:PFMR}
 \end{minipage}
\end{figure*}
\subsection{L3 Miss Rate}

The L3 Miss Rate (Eq.~\ref{eq:L3MR}) of the SpMV kernel behaves analogously to the L2 Miss
Rate, only shifted by the increased cache size, as shown in
Figure~\ref{fig:L3MR}. Just as with the L2 Miss Rate, L3 Miss Rate for
FD matrices is consistently low, at about 0.1 misses per thousand
instructions. For R-MAT matrices, the L3 Miss Rate increases dramatically, to about 25
misses per thousand instructions, at the point where the problem size
exceeds the capacity of the L3 cache.



The L3 Miss Rate data parallels the observation made in the previous
section of the pattern of fetching data from the L2 cache. For the FD
matrices, the prefetcher successfully loads data into the L2
cache. Because the SpMV kernel nearly always finds the data it needs
in the L2 cache, it rarely needs to access the L3 cache. Consequently,
the L3 Miss Rate (which is normalized by the number of instructions,
not the number of L3 accesses) is low (good) for FD matrices. For the
R-MAT matrices, the L3 Miss Rate follows the same pattern as the L2
Miss Rate: once the problem size no longer fits in cache, the miss
rate increases sharply.

For both types of matrices, the L3 Miss Rate approaches the L2 Miss
Rate. In the case of the largest matrices (for which only a trivial
portion of the problem fits in the L3 cache) nearly every L2 miss is
followed by an L3 miss. This implies that when the SpMV kernel looks
for data in the L2 cache and misses, it consistently misses in the L3
cache as well and must access DRAM to retrieve the data. This shows
that the L3 cache rarely contains relevant data and that L3 cache
accesses merely waste compute cycles.

\subsection{L2 Stall Cycles \& Prefetch Miss Rate}

L2 Stall Cycles (Eq.~\ref{eq:L2SC}) in Figure~\ref{fig:L2SC} shows the percentage of total
cycles that are wasted due to L2 cache misses. High L2 Stall Cycles
indicates congestion in the memory hierarchy and shows that the SpMV
kernel is performing poorly. High L2 and L3 Miss Rates both contribute
to L2 Stall Cycles. An L2 miss leads to an L3 access, causing the
processor to stall for several cycles. An L3 miss leads to a DRAM
access, causing the processor to stall for many more cycles. In the
case where most L2 misses are followed by L3 misses (which happens for
both types of matrices), an L2 miss is extremely costly.


For FD matrices, L2 Stall Cycles is less than one percent for problem
sizes that fit in the L3 cache. On larger matrices, L2 Stall Cycles
increases, especially in the parallel case. Although the L3 Miss Rate
is still low for these large matrices, high Prefetch Miss Rate (as
shown in Figure~\ref{fig:PFMR}, from Eq.~\ref{eq:PFMR}) congests
memory and causes stalls. A high Prefetch Miss Rate shows that the
prefetcher is successfully anticipating data needs and making the SpMV
kernel more efficient. However, the tradeoff is additional traffic to DRAM, which increases the L2 Stall Cycles resulting
from an L3 miss.\footnote{For small problems that fit entirely in the
  L2 cache, the SpMV kernel does not miss in L2, so the Prefetch Miss
  Rate does not factor into L2 Stall Cycles.} In the serial case,
prefetcher activity does not stall the processor significantly, but in
the threaded case, each core has its own prefetcher, so prefetcher
activity causes congestion and stall cycles.

For R-MAT matrices, L2 Stall Cycles begins to increase at the point
when the problem size exceeds the capacity of the L2 cache (where the L2
Miss Rate increases). The effect of the L2 Miss Rate on L2 Stall Cycles
is exacerbated by the L3 Miss Rate, particularly once the problem size
exceeds the capacity of the L3 cache (where the L3 Miss Rate jumps). As
more requests go to DRAM, the DRAM quickly reaches a
bottleneck in the rate at which it can fulfill requests. The Prefetch
Rate indicates that the congestion in DRAM overwhelms the prefetcher,
causing the prefetcher to be shut off. Once the DRAM bottleneck is
reached, the L2 Stall Cycles plateaus around 70\%. This means that,
for the largest R-MAT matrices, at most 30\% of cycles are used for
computation of the SpMV kernel, and the rest are wasted stalling.






\subsection{GFLOPS}
\label{sec:gflops}

\begin{figure*}[t]
  \centering
	\begin{subfigure}[b]{.98\columnwidth}
        	\centering       
        	\includegraphics[width=1\textwidth]{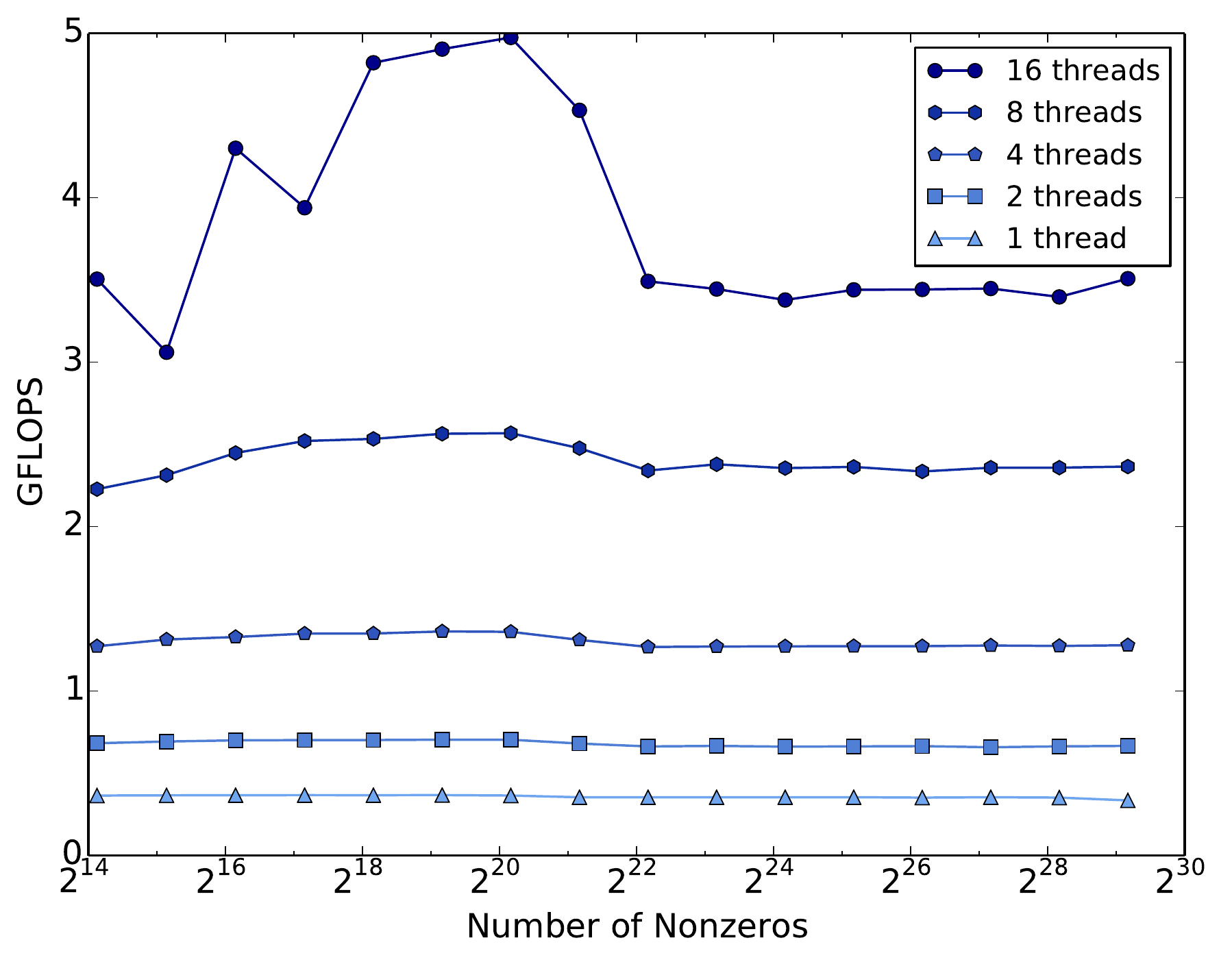} 
        	\caption{Performance in GFLOPS for FD matrices.}
        	\label{fig:gflopsFD9}
	\end{subfigure}
	\hspace{\columnsep}
	\begin{subfigure}[b]{.98\columnwidth}
        	\centering       
        	\includegraphics[width=1\textwidth]{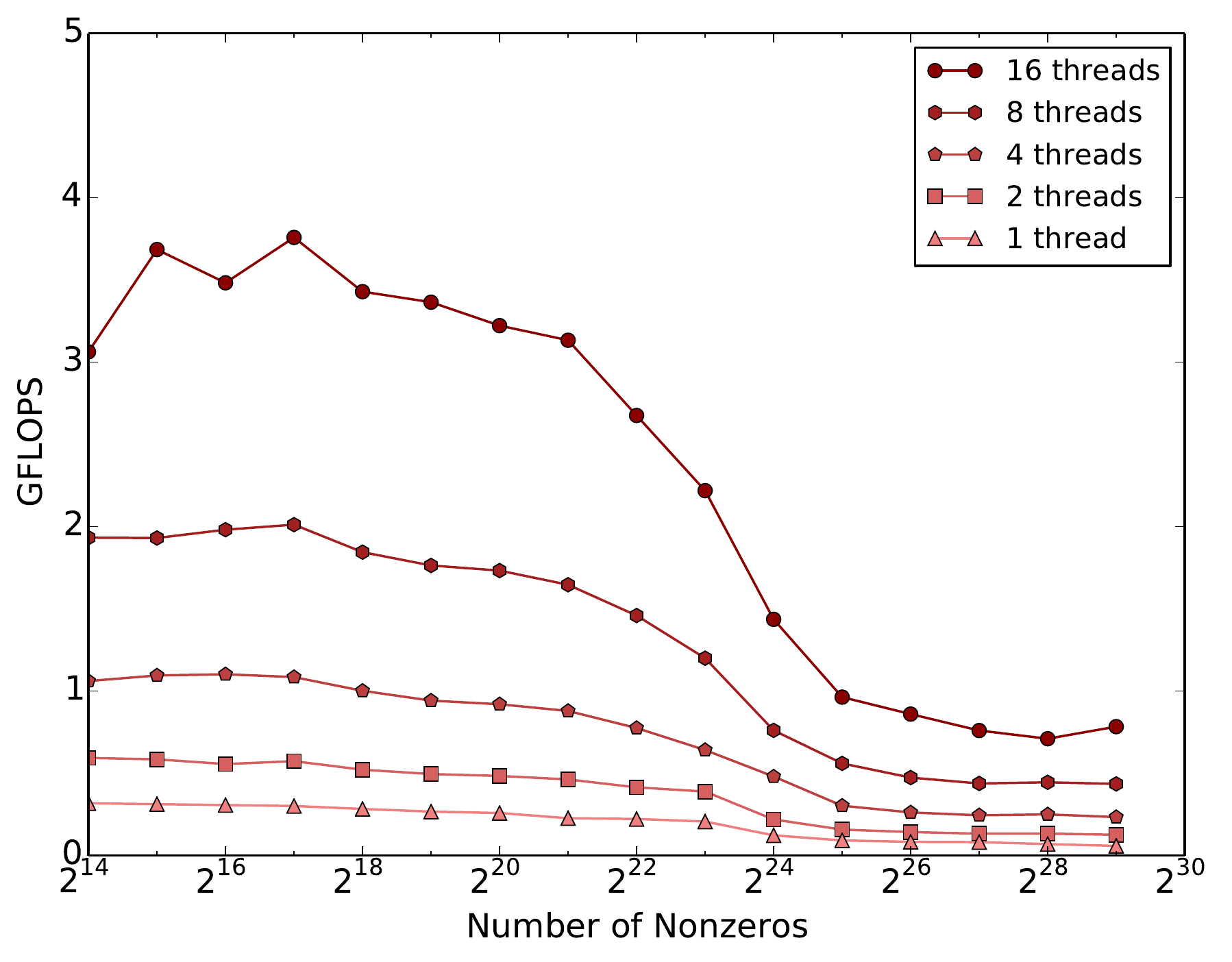} 
        	\caption{Performance in GFLOPS for R-MAT matrices.}
       		\label{fig:gflopsRMAT}
	\end{subfigure}
        \caption{Performance in GFLOPS for the SpMV kernel across matrix size for 1, 2, 4, 8, and 16 threads.}
        \label{fig:gflops}
\end{figure*}

Figure~\ref{fig:gflopsFD9} shows the performance in GFLOPS (Eq.~\ref{eq:GFLOPS}) of the SpMV kernel on FD matrices. The data
indicates that doubling the number of threads approximately doubles
the GFLOPS. This shows that the SpMV kernel is able take advantage of
the resources provided by additional threads. The performance in
GFLOPS remains constant across the number of nonzeros, indicating that
the SpMV kernel generally scales well with matrix size, with the
exception of 16 threads. The behavior on 16 threads differs due to a
bottleneck in DRAM accesses. The prefetcher causes a large amount of
DRAM activity, which, although beneficial overall, slows demand
requests and causes the SpMV kernel to stall.

Similar to the FD matrices, the SpMV kernel with R-MAT matrices is
able to utilize the resources provided by additional threads, as shown
in Figure~\ref{fig:gflopsRMAT}. Unlike for FD matrices, increasing
R-MAT matrix size causes GFLOPS to decline. In particular, when the
problem size exceeds the capacity of the L3 cache, the SpMV kernel
begins to be limited by a bottleneck in DRAM accesses, causing a
performance drop.

\section{Conclusions}
\label{sec:conclusion}


Matrix structure plays a strong role in the performance of the SpMV
kernel. Structured matrices allow the SpMV kernel to take advantage of
prefetching and caches to achieve good performance. Unstructured
matrices, however, inhibit the utilization of these architectural
features and cause performance to suffer.

For FD matrices, the regular matrix structure leads to a regular data
access pattern that is easy for the architecture to handle. This leads
to low L2 and L3 cache miss rates, high prefetcher effectiveness, and
low L2 Stall Cycles. The result is high GFLOPS, which shows high
performance of the SpMV kernel with FD matrices.

For R-MAT matrices, the random matrix structure creates an
irregular data access pattern, which is difficult for the architecture
to anticipate. Even though the matrix itself is stored in a structured
format (like the FD matrices), the erratic accesses to $\vec{x}$ cause
high L2 and L3 cache miss rates, prefetcher failure, and high L2 Stall
Cycles. The result is low GFLOPS, which shows low performance of the
SpMV kernel. More specifically, the performance in GFLOPS of the SpMV
kernel with large R-MAT matrices is only 20\% of the GFLOPS SpMV achieves
with large FD matrices.

There are several potential solutions to mitigate the problem of poor SpMV performance on unstructured, sparse
matrices. One possibility is
to bypass the L3 cache for larger problems. As the L2 and L3 miss
rates show, the L3 cache rarely contains data useful to the SpMV kernel, even in
the case of FD matrices, so performance would be improved by not
accessing the L3 cache at all. This would save power and time on standard
hardware, and potentially save chip space on specialized hardware.
The prefetcher would also perform better if it were exclusively
fetching matrix data, since it is stored in a dense format, rather than
also fetching portions of $\vec{x}$. In conjunction with this more refined
prefetcher activity, most of the cache could be used to store the
values from $\vec{x}$ (rather than extra matrix data, which the
prefetcher could retrieve easily).  A more sophisticated improvement to the
prefetcher would be to give it a more intelligent strategy to predict
data requirements. The prefetcher is able to anticipate memory accesses that are sequential, but is currently unable to predict non-sequential data needs. If the
prefetcher were able to predict the non-sequential data, the structure of the
matrix would not matter.  In general, these improvements would require
allowing the kernel more control over the architecture and tailoring
the architecture usage for the particular problem of performing SpMV
on unstructured, sparse matrices.


%

\section*{Acknowledgments}

The authors want to thank the DARPA MTO for support
of the PAKCK study. We thank Albert Reuther his
contributions.

\IEEEtriggeratref{11}


\bibliographystyle{IEEEtran}
\bibliography{PAKCK_Sparse_HPEC2014}

%



\end{document}